\documentclass{vgtc}                          

\graphicspath{{figures/}{pictures/}{images/}{./}} 

\usepackage{times}                     

\usepackage{tabu}                      
\usepackage{booktabs}                  
\usepackage{lipsum}                    
\usepackage{mwe}                       
\usepackage[normalem]{ulem}

\usepackage{mathptmx}                  

\onlineid{0}

\vgtccategory{Research}

\vgtcinsertpkg

\title{
\vspace{-2em} 
\normalsize\textsf{This survey has been expanded and published at \textit{Findings of EMNLP 2025}. Please cite the version of record:}\\
\normalsize\href{https://aclanthology.org/2025.findings-emnlp.506/}{\textit``{Modeling, Evaluating, and Embodying Personality in LLMs: A Survey}''}\\[2em]
\LARGE Integrating Personality into Digital Humans: A Review of LLM-Driven Approaches for Virtual Reality
}

\author{
Iago A. Brito\thanks{e-mail: iagoalves@discente.ufg.br}\\ %
     \parbox{3.0in}{\scriptsize \centering Advanced Knowledge Center in Immersive Technologies\\ Federal University of Goiás}
\and Julia S. Dollis\thanks{e-mail: juliadollis@discente.ufg.br}\\ %
     \parbox{3.0in}{\scriptsize \centering Advanced Knowledge Center in Immersive Technologies\\ Federal University of Goiás}
\and Fernanda B. F\"arber\thanks{e-mail: fernandabufon@discente.ufg.br}\\ %
     \parbox{3.0in}{\scriptsize \centering Advanced Knowledge Center in Immersive Technologies\\ Federal University of Goiás}
\and Pedro S. F. B. Ribeiro\thanks{e-mail: schindler@discente.ufg.br}\\ %
     \parbox{3.0in}{\scriptsize \centering Advanced Knowledge Center in Immersive Technologies\\ Federal University of Goiás}
\and Rafael T. Sousa\thanks{e-mail: rafaelsousa@ufmt.br}\\ %
     \parbox{3.0in}{\scriptsize \centering Advanced Knowledge Center in Immersive Technologies\\ Federal University of Mato Grosso}
\and Arlindo R. Galv\~ao Filho\thanks{e-mail: arlindogalvao@ufg.br}\\ %
     \parbox{3.0in}{\scriptsize \centering Advanced Knowledge Center in Immersive Technologies\\ Federal University of Goiás}
 }

\abstract{
The integration of large language models (LLMs) into virtual reality (VR) environments has opened new pathways for creating more immersive and interactive digital humans. By leveraging the generative capabilities of LLMs alongside multimodal outputs such as facial expressions and gestures, virtual agents can simulate human-like personalities and emotions, fostering richer and more engaging user experiences. This paper provides a comprehensive review of methods for enabling digital humans to adopt nuanced personality traits, exploring approaches such as zero-shot, few-shot, and fine-tuning. Additionally, it highlights the challenges of integrating LLM-driven personality traits into VR, including computational demands, latency issues, and the lack of standardized evaluation frameworks for multimodal interactions. By addressing these gaps, this work lays a foundation for advancing applications in education, therapy, and gaming, while fostering interdisciplinary collaboration to redefine human-computer interaction in VR.
} 

\keywords{Digital humans, virtual reality, large language models.}

\begin{document}


\firstsection{Introduction}

\maketitle

Virtual reality (VR) technology has emerged as a powerful medium to create and replicate real-world scenarios in digital spaces, offering unparalleled opportunities for simulation and interaction \cite{pedagogic_agent, mihai_digitaltwin}. Recent advancements in artificial intelligence have further enhanced VR environments by enabling more dynamic scenarios, such as generating text-based descriptions for 360-degree VR scenes \cite{elisasvr} and user attention-guidance in virtual tours around real locations using multi-modal models \cite{paulosvr}.  Complementing these advancements, studies on computer-driven characters with human-like characteristics, referred to as digital humans, highlight their significant role in enhancing user engagement within VR environments by providing realistic facial expressions, responsive gestures, and natural conversational abilities, mimicking humans not only in their visual design but also in their personality, decreasing the gap between virtual and real-world interactions \cite{hube2024study, thomas2022investigatinghow}.

The potential applications of digital humans in VR environment passes through  a variety of fields. In educational settings, the use of virtual pedagogic agents with human-like characteristics demonstrate gains in students’ learning experience and engagement \cite{pedagogic_agent}, while in therapeutic context, the use of digital humans combined with therapy techniques demonstrates gains in reducing symptoms of borderline personality disorder (BPD) \cite{Falconer123}. These possibilities underscore the importance of further research into designing, implementing, and evaluating such agents in VR. 

In parallel, the field of natural language processing (NLP) has witnessed transformative advancements with the rise of large language models (LLMs), such as GPT \cite{brown2020languagemodelsfewshotlearners} and LLaMA \cite{touvron2023llamaopenefficientfoundation} family of models. Trained on vast datasets containing trillions of words, these models demonstrate emergent abilities (i.e., capabilities that arise without explicit training for specific tasks), like the capacity to simulate human personality and emotions in real-time conversations, providing an unprecedented means for human-computer interaction \cite{shao-etal-2023-character, kroczek2024influencepersonaconversationaltask, tseng-etal-2024-two}. These capabilities represent a critical step forward in creating digital agents that can think and respond in ways that feel authentically human.

Despite their potential, most applications of these generative models remain confined to chat-based interfaces. Although such systems offer interactive benefits, they do not simulate real-world communication as effectively as VR technology. The integration of NLP-driven digital humans into VR offers a transformative alternative: enabling face-to-face interactions where digital agents can express themselves through both verbal and non-verbal cues, such as facial expressions, gestures, and personality-driven behavior \cite{taylor2023rise, maslych2025takeawaysapplyingllmcapabilities}. This multimodal interaction is closer to real human-to-human communication, offering users a more engaging and satisfying experience than chat interfaces.

Furthermore, studies have shown that digital humans personality directly influences how users perceive and interact with the system, impacting key metrics like immersion, trust, and engagement \cite{SACCO2014359, kroczek2024influencepersonaconversationaltask}. By embedding these agents with the ability to convey personality traits and human-like expressions, VR can promote deeper connections and more meaningful interactions, paving the way for future advancements in virtual communication.

This work provides a comprehensive review of methods for enabling digital humans to adopt personas and identities, focusing on approaches developed both within and outside the VR context and exploring their adaptation to immersive environments. It examines how large language models can simulate human-like traits, combining verbal and non-verbal communication to enhance immersion and realism. The paper also highlights the challenges of integrating LLM-driven personality traits into VR, including computational demands, latency issues, and the lack of standardized evaluation frameworks for multimodal interactions. By addressing these gaps, this work lays the foundation for advancing applications in education, therapy, and interactive simulations, fostering more engaging and human-like virtual interactions.

\section{Digital Humans}

Digital humans (DH) have become a focal point of research alongside the rapid expansion of VR technology. These entities are computer-generated representations of human beings, designed to replicate physical appearances, cognitive functions, and emotional responses in digital environments \cite{lin2024human, HU201164}. By simulating human behavior across multiple dimensions, digital humans aim to bridge the gap between artificial systems and natural interactions, unlocking new possibilities for immersive experiences.

Despite their potential, DH often fail to exhibit the nuanced personalities and emotional depth necessary to promote natural and engaging interactions. This absence of relatable human-like traits often results in interactions that feel mechanical and impersonal, making it difficult for users to form meaningful connections with these agents \cite{WESTER2024100072, sonlu2024effectsembodimentpersonalityexpression}. By designing these agents with more human-like behavior and expressions, these systems can offer greater relatability, which enhances the overall interaction experience and promotes stronger user engagement.

By aligning digital humans behaviors with user expectations and situational needs, these personality-driven agents promote a deeper sense of connection, making interactions more meaningful and enjoyable \cite{SACCO2014359}. Digital humans with empathetic personality demonstrates improvements in learning experiences by creating a supportive environment where users feel understood and motivated, ultimately increasing their willingness to engage and learn \cite{sonlu2024effectsembodimentpersonalityexpression}.

In virtual reality environments, agents with distinct and well-defined personalities, conveyed through both textual and visual features, play a critical role in improving user engagement and immersion. By emulating human-like behaviors and expressions, these agents enable interactions that are both dynamic and natural \cite{sonlu2024effectsembodimentpersonalityexpression, pan2024ellmatembodiedllmagentsupporting}. Moreover, by embedding distinct personality traits into virtual reality scenarios, they play a pivotal role in enhancing both the realism and the functional effectiveness of digital humans.

\subsection{Personality-driven Agents}

Personality conditioning refers to the process of equipping virtual agents with human-like traits and behaviors to enhance the quality of their interactions with users \cite{SENDRES1995477}. Early approaches to personality conditioning utilized rule-based systems and decision trees to simulate behavior. These systems mapped specific inputs to predefined outputs, functioning within a limited framework of programmed responses \cite{1966}. While innovative for their time, these methods were inherently rigid, lacking the adaptability to handle the unpredictability of human interactions. This rigidity often resulted in repetitive, unnatural conversations that failed to engage users meaningfully or address the complexities of real-world scenarios.

The integration of verbal communication with non-verbal cues has proven essential in enhancing interactions between humans and virtual agents, creating experiences that feel more natural and immersive \cite{Cassell_2001}. By combining elements such as facial expressions, gestures, and tone of voice, these systems enable dynamic and responsive interactions. This multimodal approach not only enriches conversational content but also improves conversational flow, fostering mutual understanding and creating a stronger sense of presence in virtual environments.

Building on these foundational principles, recent advancements in embodied virtual agents have pushed the boundaries of natural and effective interactions. For example, \cite{2015} developed a system designed to enhance Willingness to Communicate (WTC) in second-language learners. By incorporating context-dependent domain knowledge and user intent detection, the agent managed conversational flow in a way that was both semantic and user-centric. Learners reported increased confidence and a greater desire to communicate in a foreign language after interacting with the system, demonstrating its potential for language acquisition.

However, even with these advancements, capturing the complexity of human personality within these systems remained a challenge. The absence of subtle, context-aware responses can lead to interactions that, while functional, lack the depth and authenticity required for truly lifelike experiences. This highlighted the need for more advanced approaches, which LLMs later addressed by allowing more nuanced and flexible personality modeling capabilities.

\section{Large Language Models (LLMs)}

The field of natural language processing (NLP) has undergone rapid transformations since the introduction of the Transformer architecture \cite{vaswani2023attentionneed}. This breakthrough has fundamentally reshaped NLP, leading to the proliferation of numerous language models, particularly decoder-based architectures, which have revolutionized text generation and paved the way for large language models (LLMs). These advanced neural networks redefine the boundaries of machine understanding and language generation, leveraging extensive datasets comprising billions or even trillions of words from diverse domains to maximize linguistic coverage and generalization \cite{brown2020languagemodelsfewshotlearners}.

The unprecedented scale of LLMs, characterized by vast numbers of parameters, enables them to capture intricate linguistic patterns, including semantic relationships, syntactic structures, and long-range dependencies \cite{touvron2023llamaopenefficientfoundation, brown2020languagemodelsfewshotlearners}. This capacity allows them to perform complex language-related tasks, such as advanced reasoning and contextual understanding. In particular, LLMs demonstrate emergent abilities, performing tasks that are not explicitly programmed but naturally arise from their scale and the complexity of their training process \cite{wei2022emergentabilitieslargelanguage, kaplan2020scalinglawsneurallanguage}.

The development of large language models generally involves two primary stages: pre-training and fine-tuning. The pre-training phase involves exposing the model to extensive volumes of unlabeled general knowledge text data through a language modeling task, most commonly framed as the next token prediction in a sequence \cite{brown2020languagemodelsfewshotlearners}. Then, the resulting model passes through the fine-tuning phase, which refines the model’s capabilities to perform a more specific set of tasks. This process leverages the pre-trained weights, which encode general knowledge about language, morphology, and syntax, and adjusts them using labeled datasets in a supervised learning framework \cite{dong-etal-2024-abilities}. This step, commonly referred to as Supervised Fine-Tuning (SFT), aligns the model’s performance with the requirements of particular applications, significantly enhancing its ability to tackle domain-specific tasks \cite{mecklenburg2024injectingnewknowledgelarge}.

While fine-tuning significantly enhances a model's domain-specific performance, it often requires access to a large amount of labeled data, which presents a substantial barrier, especially when dealing with specialized domains. This challenge has motivated research into alternative approaches to task adaptation, like In-Context Learning (ICL) \cite{li-2023-practical}, which allows learning different tasks by textual instructions at inference time using techniques such as zero-shot and few-shot learning \cite{brown2020languagemodelsfewshotlearners, touvron2023llamaopenefficientfoundation}.

This shift of paradigm enables LLMs to execute complex tasks without requiring updates to their parameters or access to extensive labeled datasets in the target domain. Instead, prompts function as contextual frameworks containing either examples of the target task or explicit natural language instructions.

Such adaptability is instrumental in advancing large language models, fostering the development of systems that can dynamically learn, adjust to new contexts, and execute tasks requiring intricate understanding. Among these advancements, a particularly noteworthy avenue is the simulation of human-like behaviors, such as the ability to exhibit nuanced linguistic and contextual adaptations. Although emergent abilities have significantly enhanced natural language understanding and generation, their implications for enabling coherent and consistent personality modeling in virtual agents present a promising yet underexplored research frontier.

\subsection{Modelling Personality Through LLMs}

Building on the concept of emergent abilities in large language models, the task of represent human personality into virtual agents represents an exciting and emergent domain. Personality modeling seeks to create systems that not only respond contextually but also reflect consistent psychological characteristics, enhancing user engagement and creating more lifelike interactions \cite{SENDRES1995477}.

Research into personality modeling within LLMs has primarily focused on understanding the intrinsic personality traits of these models and developing methods to control or induce specific traits. Studies such as \cite{pan2023llmspossesspersonalitymaking} and \cite{serapiogarcía2023personalitytraitslargelanguage} have analyzed the underlying characteristics of LLMs using established psychological frameworks, such as the MBTI \cite{myers1988myers} and Big Five \cite{bigfive} personality traits, highlighting the inherent tendencies of LLMs emerged from the training data and model architecture, and exploring how these tendencies might align or deviate from human-like personality expressions.

Personality modeling in Large Language Models (LLMs) predominantly employs three primary approaches: zero-shot learning, few-shot learning, and fine-tuning.

\begin{enumerate}
    \item \textbf{Zero-Shot Learning:}

By crafting precise instructions that implicitly convey desired personality traits, these methods guide the model to generate responses aligned with specific personality characteristics. For instance, the approach described in \cite{jiang-etal-2024-personallm} uses descriptive prompts that can elicit specific personality traits by framing the context in a way that enables the model to naturally embody the intended personality characteristics. 
    \item \textbf{Few-Shot Learning:}

This method involves providing the model with a limited number of carefully selected examples that demonstrates the desired personality characteristics, thus improving the ability of the model to generalize these traits in subsequent interactions. An illustrative example is the study  \cite{zhu2024personalityalignmentlargelanguage} where a limited number of carefully crafted examples from the training data (e.g., responses from the IPIP-NEO-120 questionnaire) are provided within the prompt. 
    \item \textbf{Fine-Tuning:}

This supervised learning process modifies the pre-trained weights using labeled datasets that reflect the desired personality characteristics, thereby enhancing the model’s ability to consistently exhibit these traits across diverse tasks. An example of this approach is \cite{shao-etal-2023-character} which integrates historical context and behavioral traits into the training process to create historically grounded personas. By refining the model’s internal representations, fine-tuning ensures that the LLM not only understands general language constructs but also reliably manifests the specified personality attributes in its outputs. 

\end{enumerate}

Despite significant advancements in personality modeling using large language models (LLMs), much of the research remains centered on text-based chatbot interactions. For example, studies such as \cite{huang2024orcaenhancingroleplayingabilities} and \cite{ran2024capturingmindsjustwords} delve into LLM-driven personalities within text or role-playing contexts. Similarly, \cite{klinkert2024drivinggenerativeagentspersonality} investigates personality applications in non-player characters (NPCs) within gaming environments. However, these works often neglect the unique challenges of virtual reality, where embodied agents interact with users in immersive, real-time scenarios. Addressing these challenges requires integrating multimodal communication, ensuring responsiveness, and creating an engaging sense of presence, areas that remain underexplored in the current literature.

Virtual reality introduces distinct opportunities by enabling interactions that are richer and more immersive than text-based systems. The use of gestures, facial expressions, and gaze enhances the expression of personality, making interactions more dynamic and engaging. As highlighted in \cite{kroczek2024influencepersonaconversationaltask}, such interactions emulate natural human communication, fostering deeper social connections and user engagement. While these advantages highlight VR’s potential, they also underscore the need for research focused on optimizing personality-driven digital agents for this medium, ensuring that non-verbal communication aligns seamlessly with verbal outputs to create truly lifelike interactions.

The ability to express personality through both text and non-verbal cues, such as gestures and facial expressions, is crucial for enhancing user experience, underscoring the need for further research into seamlessly integrating these expressive modalities into VR environments. While some studies explore multimodal interactions, they often fall short of fully addressing their application in VR. For instance, \cite{Cai_2024_CVPR, cherakara2023furchatembodiedconversationalagent, 99} investigate simulations where agents communicate through speech and gestures or delve into the role of facial expressions, yet they either briefly mention the potential integration of VR or omit it entirely.

Despite the scarcity of research on incorporating LLM-driven personality traits into digital humans in VR, a few studies have begun exploring this convergence. For example, \cite{hasan_sapienaffective} employs multimodal technologies to enhance human-computer interaction, demonstrating the use of LLMs in real-time, open-domain conversations across multiple languages while incorporating speech synthesis and facial expressions. By aligning verbal communication with non-verbal emotional cues, this work illustrates the feasibility of combining LLMs and multimodal systems to create virtual agents with heightened contextual and emotional responsiveness. Such implementations highlight the potential of these technologies to foster more engaging and lifelike interactions in VR environments.

This absence of literature underscores the necessity for more comprehensive studies that bridge these two fields, paving the way for better and more engaging ways to interact with advanced virtual agents with rich personalities.

\section{Evaluating Digital Humans Personality Traits}

As digital humans become increasingly integrated into daily life, their behavior and impact on user interactions have become areas of growing interest \cite{yang2024makesmodellowempathywarmth, noever2023aitexttobehaviorstudysteerability}. In the field of psychology, personality traits are defined as enduring patterns of thoughts, feelings, and behaviors that exhibit consistency in time and situations, underscoring individual differences along a set of basic dimensions, reflecting stability and predictability in behavior \cite{diener2019personality}.
 
The systematic study of personality traits has a long history in psychology, aiming to understand and categorize individual differences \cite{cattell1977scientific}. Trait theories provide structured frameworks to explore how people think, feel, and behave in consistent ways over time and across situations. Broad frameworks, like the Big Five model \cite{de2000big}, focus on overarching dimensions of personality, organizing personality into five broad dimensions—openness, conscientiousness, extroversion, agreeableness, and neuroticism, capturing fundamental aspects of how individuals engage with the world. More detailed approaches, such as the Sixteen Personality Factors (16PF) \cite{cattell2008sixteen}, break down personality into 16 specific traits, such as emotional stability, social courage, and sensitivity.  

Collectively, these methods provide structured tools to explore the complexity of personality, fostering a deeper understanding of the stable characteristics that shape individual differences. Furthermore, research has consistently shown a strong correlation between personality and language use \cite{hirsh2009personality, pennebaker1999linguistic, pennebaker2001patterns, lee2007relations}. This relationship has enabled the evaluation of personality traits through text analysis, allowing  researchers to assess individual characteristics by examining the frequency and patterns of words and expressions used.


Although foundational theories commonly applied to human beings have inspired efforts to evaluate personality traits in LLMs \cite{jiang-etal-2024-personallm, hagendorff2023machine, wen2024selfassessmentexhibitionrecognitionreview, vu2025psychadapteradaptingllmtransformers}, it remains a complex and nuanced challenge, particularly in digital humans, as it involves a highly subjective context-dependent and non-standardized domain. Existing methods for evaluating LLM personality traits include human evaluations \cite{feng2024sampleefficienthumanevaluationlarge, guzman-etal-2015-humans}, LLM-as-judge approaches \cite{li2024llmsasjudgescomprehensivesurveyllmbased, li2025generationjudgmentopportunitieschallenges}, and adaptations of self-assessment personality tests originally designed for humans \cite{song2023largelanguagemodelsdeveloped, karra2023estimatingpersonalitywhiteboxlanguage}. 

These approaches have primarily focused on text-based interactions, examining whether LLMs exhibit consistent and stable personality traits \cite{song2024identifyingmultiplepersonalitieslarge, song2023largelanguagemodelsdeveloped}, exploring methodologies for personality assessment \cite{lee2024llmsdistinctconsistentpersonality, zou2024llmselfreportevaluatingvalidity}, investigating the impact of personality on model safety \cite{zhang2024betterangelsmachinepersonality}, and tailoring personality traits to specific use cases \cite{li2024big5chatshapingllmpersonalities, huang2024designingllmagentspersonalitiespsychometric, vu2025psychadapteradaptingllmtransformers}. These efforts have established a foundation for evaluating personality in text-based agents but fall short when considering the multimodal nature of embodied conversation agents. Due to its complex layers of interaction and connection, we are introduced to challenges that existing evaluation methods, which primarily focus on text-based outputs, are not equipped to address. In that way, significant gaps remain in the standardization and application of multimodal evaluation frameworks.

To address this gap, it is crucial to critically analyze existing methods for assessing personality traits in LLMs and consider how these approaches can be adapted for VR contexts and embodied conversational agents. The following sections examine qualitative and quantitative evaluation methods for LLMs-driven personality traits, highlighting their strengths, limitations, and potential applicability to computer-driven characters.

\subsection{Qualitative Evaluation}
As previously discussed, evaluating the personality traits of digital humans involves complex, nuanced, and non-standardized methods. Among these, qualitative evaluation methods are widely used, particularly in LLM-based conversational models \cite{li2024llmsasjudgescomprehensivesurveyllmbased}. These methods are primarily subjective, relying on human (or, in some cases, LLM) interpretation to assess an agent's performance, coherence, and adherence to specific personality traits. While qualitative evaluations are valuable for understanding how users perceive an agent's behavior and personality—critical for improving user-agent relationships—they lack standardization \cite{chen-etal-2024-humans}. To address the limitations of purely human or purely LLM-based evaluations, many studies adopt a hybrid approach, combining the strengths of both methods \cite{jiang-etal-2024-personallm}. However, the inherent subjectivity of these methods means that experiences and interpretations can vary significantly from person to person (or model to model, in the case of LLM-as-judge evaluations).

Human evaluation has historically been one of the most widely used qualitative methods in research, particularly for assessing alignment in LLM-based models. This approach evaluates whether an LLM's behavior aligns with human intentions, ethical values, and predefined goals \cite{abeysinghe2024challengesevaluatingllmapplications}. Beyond alignment, it serves as a critical tool for evaluating the personality traits and behavioral consistency of LLMs \cite{vu2025psychadapteradaptingllmtransformers}. This method typically involves interactive tests, where evaluators engage directly with the agent to determine whether it effectively exhibits the desired personality traits and meets user expectations \cite{jiang-etal-2024-personallm}. Human evaluation can be conducted from multiple perspectives, enabling a comprehensive analysis of various dimensions of the agent’s behavior. These distinct angles of evaluation provide deeper insights into the agent’s overall performance and its adherence to intended personality traits and behavioral standards.

Human evaluation has been widely used to assess how personality traits influence user perceptions of virtual agents during conversational tasks, offering valuable insights into social interactions. Research in this area has explored virtual agents designed to exhibit extroverted or introverted characteristics, expressed through variations in tone, communicative style, and level of engagement \cite{kroczek2024influencepersonaconversationaltask}. User assessments based on factors such as warmth, empathy, and realism have demonstrated the significant role of personality traits in shaping perceptions and interactions in virtual environments.

However, human evaluation as a methodology presents notable limitations. Its subjective nature introduces variability, as individual evaluators may interpret and score interactions differently, leading to inconsistencies, particularly in studies involving multiple evaluators with diverse perspectives. This variability can undermine the reliability of results and complicate cross-study comparisons. Furthermore, the effectiveness of human evaluation as a primary approach to assess the personality traits of digital humans remains a topic of ongoing debate within the research community \cite{clark2021thatshumangoldevaluating}. These challenges highlight the need for standardized evaluation protocols and objective metrics to improve consistency and comparability between studies.

Another qualitative evaluation method involves the use of the LLM-as-judge approach. This emerging framework employs LLMs to evaluate the personality and behavior of other LLMs and conversational agents by simulating human judgment \cite{li2024llmsasjudgescomprehensivesurveyllmbased}. The LLM-as-judge methodology has gained traction due to its scalability and ability to automate the evaluation process \cite{dong2024llmpersonalizedjudge, zhu2023judgelmfinetunedlargelanguage}, reducing the dependence on human evaluators, which can be time-consuming and costly.

In this approach, an LLM is provided with a prompt containing the text to be assessed along with clear evaluation criteria. For example, the criteria might involve a question designed to assess specific personality traits of the tested LLM. Based on the prompt, the evaluating LLM assigns scores or labels to the text according to the predefined criteria. This method has been widely employed to evaluate conversational agents, models, and chatbots across various domains \cite{wei2024systematicevaluationllmasajudgellm}, including the context of this research \cite{tseng-etal-2024-two}.

Although the LLM-as-judge approach offers notable benefits, such as increased efficiency and reduced costs for evaluating LLM personality traits, it presents several challenges, including inherent biases in the models, sensitivity to different prompts, limitations specific to individual models, and concerns regarding the consistency and reliability of the evaluations \cite{chen2024humansllmsjudgestudy, zheng2023judgingllmasajudgemtbenchchatbot}. Addressing these issues is essential to enhance the accuracy and fairness of this methodology, highlighting the need for further refinement and the establishment of standardized evaluation frameworks.

\subsection{Quantitative Evaluation}
Quantitative evaluation methods have become an essential tool for assessing personality traits in LLM-controlled agents, offering a structured and objective approach. These methods focus on analyzing the behavior of language models based on outputs such as responses to questions, storytelling, and other text-based interactions. 

Common approaches, such as personality assesments \cite{weiner2017handbook} and word frequency analysis \cite{tausczik2010psychological}, aim to provide standardization and reproducibility, addressing the variability and subjectivity often associated with qualitative assessments. However, they still face challenges, including ongoing debates about their reliability in accurately evaluating LLM personality traits \cite{gupta-etal-2024-self}. Despite these limitations, quantitative methods provide a valuable foundation for comparing models and identifying consistent patterns in personality expression, making them indispensable for evaluating digital agents in a systematic manner.

Among quantitative evaluation methods, personality assessments originally developed for the psychological evaluation of humans \cite{pilbeam2012psychometric, weiner2017handbook} have been extensively employed to analyze a wide range of personality characteristics in LLM-driven agents. These assessments serve as systematic tools designed to measure diverse dimensions of personality, encompassing traits, behaviors, motivations, and interpersonal dynamics. Instruments such as the Big Five Inventory (BFI) \cite{john1991big}, the International Personality Item Pool (IPIP) \cite{goldberg2006international}, and the NEO Personality Inventory (NEO-PI-3) \cite{mccrae2005neo} are among the most frequently utilized for evaluating both human subjects and virtual agents.

In these structured assessments, individuals or LLMs provide evaluations of behaviors, thoughts, and feelings by responding to a series of predetermined statements or questions. These responses, collected through formats such as Likert scales (e.g., ratings ranging from "strongly disagree" to "strongly agree") \cite{likert1932technique}, true-false options, or forced-choice formats, are scored and analyzed to produce detailed profiles of personality traits. When applied to LLMs, specific prompt instructions are often used to simulate self-reported responses, enabling the modeling of human-like personality traits \cite{jiang_2024, serapiogarcía2023personalitytraitslargelanguage, rao-etal-2023-chatgpt}.

Although widely used, these assessments present significant challenges when applied to non-human entities such as LLMs. One notable issue is the sensitivity of LLMs to multiple-choice question (MCQ) formats, as responses are heavily influenced by the order of options presented \cite{zheng2024largelanguagemodelsrobust}. Since self-reported questionnaires frequently rely on MCQs, this introduces bias into the responses generated by digital agents, compromising the reliability of the evaluations. Consequently, the administration of self-reported questionnaires to virtual agents often yields unreliable results \cite{gupta-etal-2024-self}.

To address these challenges, efforts have been made to adapt self-report questionnaires originally designed for humans into formats suitable for assessing personality traits in digital agents \cite{lee2024llmsdistinctconsistentpersonality}. One significant contribution is the Language Model Linguistic Personality Assessment (LMLPA) \cite{zheng2024lmlpalanguagemodellinguistic}. This system adapts the Big Five Inventory — a cornerstone in psychometric research \cite{john1991big} — by aligning its format with the operational capabilities of LLMs. Incorporating AI raters and linguistic prompts, the LMLPA quantitatively evaluates personality traits through the linguistic outputs of LLMs, offering a novel, domain-specific approach. Although promising, the methodology requires further empirical validation to ensure its robustness and applicability across diverse scenarios \cite{zheng2024lmlpalanguagemodellinguistic}.

Another widely employed quantitative evaluation method for assessing personality traits in LLM-based digital agents is the Linguistic Inquiry and Word Count (LIWC) software \cite{pennebaker2001linguistic}.  LIWC analyzes text by mapping words and phrases to a curated dictionary, categorizing them into dimensions reflecting psychological, emotional, and social constructs \cite{tausczik2010psychological}. Its latest version, LIWC-22, includes over 12,000 words and expressions across 117 categories, such as personal pronouns, emotion-related vocabulary, and cognitive indicators \cite{boyd2022development}. 

Widely applied in psychology \cite{tov2013detecting}, and communication research \cite{haber2015stabilityonlinelanguagefeatures}, LIWC is a valuable tool for studying personality traits in digital agents \cite{tausczik2010psychological}. In LLM-based digital agents, LIWC has been applied to assess alignment \cite{frisch2024llmagentsinteractionmeasuring}, personality retention, and emotional embedding \cite{tripto2024shiptheseuscuriouscases}. These applications highlight LIWC's ability to reveal subtle patterns and provide meaningful insights into how LLMs convey and replicate personality traits through language.

While qualitative methods are effective for analyzing personality traits in text-based agents, they are less suited for evaluating digital humans in VR environments. These agents express personality through a combination of verbal outputs and non-verbal behaviors, such as gestures, facial expressions, and vocal tone. This multimodal nature introduces complexities that demand updated evaluation methods to fully capture their dynamics. Embodied conversational agents, for example, offer a heightened sense of social presence and interaction, resembling face-to-face communication \cite{10.1145/3173574.3173863}. Solely relying on textual analysis neglects critical non-verbal cues, such as gestures and intonation, which can either reinforce or contradict verbal outputs.

The systematic assessment of personality traits in multimodal contexts remains underexplored. Current methodologies are not yet equipped to address the challenges posed by these environments, where verbal and non-verbal communication intersect. Future research should prioritize developing comprehensive evaluation frameworks that integrate facial expressions, body language, and vocal nuances with traditional textual analysis. Advancing such methodologies will enhance confidence in applying LLMs to digital humans, enabling more natural and effective interactions in VR environments.

\section{Future}

Despite rapid advancements in the field of digital humans, the use of cutting-edge AI technologies, particularly LLMs, to simulate human identity and personality in realistic, non-chat-based scenarios remains underexplored. The capabilities of generative models to mimic human behavior present a promising avenue for applications across diverse domains. For instance, LLMs can enable adaptive personalities in NPC characters, allowing non-scripted, context-driven conversations. These NPCs could maintain both short-term and long-term memory during interactions, enriching gaming experiences with more dynamic and immersive storytelling \cite{zheng2024memorynpc}. Beyond entertainment, educational applications of such agents have shown potential in improving learning outcomes, such as enhancing user satisfaction and reducing the time required to master new concepts \cite{pedagogic_agent, taylor2023rise}.

An emerging research area involves leveraging the generative capabilities of LLMs to produce text that not only conveys conversational content but also encodes directives for corresponding visual actions. By integrating emotion recognition and generation, these systems could embed instructions for digital humans to perform specific facial expressions or body movements aligned with the intended emotional tone of the interaction. This approach enables the generated text to act as both a communication medium and a control mechanism for non-verbal expressions, enhancing the realism and interactivity of virtual environments \cite{hasan_sapienaffective}. Such multimodal integration bridges the gap between verbal and non-verbal communication, creating a more seamless and human-like interaction experience.

However, while LLMs exhibit remarkable capacities for simulating and adapting human personalities, they are hindered by significant computational demands. The high inference time associated with auto-regressive generation methods translates into latency challenges, requiring substantial GPU infrastructure to ensure real-time responses \cite{xia2023flashLLM, li2024llminferenceservingsurvey}. On the other hand, recent research on small LLMs (SLLMs) suggests that these models can perform tasks previously reserved for larger models \cite{chen2024rolesmallmodelsllm}. Although still in its initial research, this field holds potential for addressing the computational challenges of LLMs, enabling the development of lightweight systems capable of delivering personality simulation and emotion-driven expressions in real time. Further research is needed to evaluate the effectiveness of SLLMs in these complex tasks, particularly in scenarios requiring both verbal and non-verbal adaptability.

\section{Conclusion}
The integration of large language models with virtual reality technologies represents a transformative step in enhancing the realism and interactivity of digital humans. By combining the verbal capabilities of LLMs with non-verbal cues such as facial expressions and gestures, these systems offer the potential to create more engaging, lifelike interactions. Despite significant advancements, this domain remains underexplored, particularly in adapting LLM-driven personality traits to human-like computer-driven characters in immersive VR environments. Addressing challenges such as computational demands, latency, and the absence of standardized evaluation frameworks is critical for realizing the full potential of these technologies.

This paper has reviewed methods for enabling digital humans to adopt nuanced personalities, highlighting the contributions of both rule-based approaches and advanced LLM techniques, including zero-shot, few-shot, and fine-tuning strategies. It also underscored the importance of multimodal integration for fostering richer user experiences and explored the opportunities and limitations of leveraging emerging technologies such as small LLMs to overcome computational constraints.

Our findings provide a foundation for further exploration, encouraging interdisciplinary collaboration between natural language processing, virtual reality, and human-computer interaction to redefine how we interact with digital humans in immersive settings.


\acknowledgments{This work has been fully funded by the project Research
and Development of Algorithms for Construction of Digital Human
Technological Components supported by the Advanced Knowledge Center in Immersive Technologies (AKCIT), with financial resources
from the PPI IoT/Manufatura 4.0 / PPI HardwareBR of the MCTI
grant number 057/2023, signed with EMBRAPII.}

\bibliographystyle{abbrv-doi}

\bibliography{template}
\end{document}